# FORENSIC ANALYSIS OF THE EXFAT ARTEFACTS

Yves Vandermeer, Nhien-An Le-Khac, Joe Carthy and Tahar Kechadi

School of Computer Science, University College Dublin, Ireland

yves.vandermeer@ucdconnect.ie, {an.lekhac, joe.carthy, tahar.kechadi}@ucd.ie

**ABSTRACT**

Although keeping some basic concepts inherited from FAT32, the exFAT file system introduces many differences, such as the new mapping scheme of directory entries. The combination of exFAT mapping scheme with the allocation of bitmap files and the use of FAT leads to new forensic possibilities. The recovery of deleted files, including fragmented ones and carving becomes more accurate compared with former forensic processes. Nowadays, the accurate and sound forensic analysis is more than ever needed, as there is a high risk of erroneous interpretation. Indeed, most of the related work in the literature on exFAT structure and forensics, is mainly based on reverse engineering research, and only few of them cover the forensic interpretation. In this paper, we propose a new methodology using of exFAT file systems features to improve the interpretation of inactive entries by using bitmap file analysis and recover the file system metadata information for carved files. Experimental results show how our approach improves the forensic interpretation accuracy.

## 1. INTRODUCTION

Today, computer forensics are more than ever used to gather evidence for all type of crimes. In addition, the file system forensics approaches have contributed to successfully solve many criminal cases. However, the evaluation of the accuracy of digital forensic tools is still a challenge. If the accuracy of computer forensics tools is part of the quality of the process, a level of understanding by the practitioner on "how" artefacts are created is required to explain "why" such content was recoverable and provide valuable contextual information to the given case.

The exFAT file system, protected by Microsoft Patent in June 2009 [1][2] has tailored to address limitation issues from previous FAT16 and FAT32 file systems by allowing larger volumes and larger files sizes and avoid some inappropriate behaviours of existing FAT and NTFS when installed on NAND storage devices. As the most recent removable storage devices are mainly NAND memory, exFAT operating system drivers try to reduce as much as possible the "write" operations to preserve device lifetime [3], especially in critical disk areas that are allocated to file systems.

Today, new SDXC memory cards make extensive use of exFAT to address new storage requirements. Several manufacturers, like car companies, already signed cooperation agreement with Microsoft to use exFAT for their embedded multimedia systems [4][5][6][7]. Only available on Microsoft Windows Operating Systems initially, exFAT is now well handled by Apple Mac OS X and quite recently by Linux distributions.

Moreover, the FAT systems are analysed by forensic practitioners for decades and extended literature covers description of FAT12, FAT16 and FAT32 structures describing how file systems artifacts can be interpreted for common events including file deletion. The FAT area plays a dual role; describing how clusters are chained and allocated. This leads to a perilous and often questionable interpretation of carved files.

Digital forensic tools have been developed with the view to investigate exFAT by using former FAT systems combined with an undocumented "magic science"; used algorithms being kept secret as competitive sale features [21]. The result is that the generated reports are sometimes inaccurate and always hermetic to practitioner's understanding on how some file status attributes were guessed [22]. The difference between a deleted and a renamed file is a good example of such issue. As an example, XWays provides for exFAT's unused set of entries a single and vague info "file moved or renamed".

Hence, the forensic analysis using a dedicated software tool needs to be built on a methodology that could make the most of exFAT specifications. By following the same methodology, practitioners will then be able to interpret the status and history of reported file in court of justice. However, file history interpretation soundness can be improved by comparing information on the cluster chaining from the FAT area with respective allocation status gathered from the bitmap allocation file. Therefore, in this paper, we aim to propose:

(i) a sound interpretation of inactive exFAT directory entries with references to the bitmap cluster allocation file.

(ii) a sound carving process based on exFAT structures, allowing to associate the remaining file system metadata with recovered file content.

(iii) a simple validation process to recover the previous content of some shortened existing files.

The rest of this paper is organised as follows: Section 2 looks at the literature survey and relevant works that have already been conducted in the field and some remaining open questions. Section 3 describes the protocol used to generate different types of files and observe associated artefacts on exFAT volumes. Section 4 describes the main components of the exFAT structure that are related to this research. Section 5 presents the proposed forensic methodology to recover deleted files and file system metadata. Finally, Section 6 concludes this research work and highlights some future work directions.

## 2. RELATED WORK

Sound File Systems forensics reference was first described by Carrier in 2005 [8] with details on the FAT chaining and other structures. Carrier proposed volume reconstruction processes, like NTFS, and facilitated by available bitmap information. However, until 2005 exFAT did not exist yet and the advantage of co-existence of a FAT and a bitmap file on same file system was not covered.

After the introduction of exFAT, SANS institute published a research work [9] about exFAT structure. In that paper, Shullich highlighted the risk of not being able to provide an accurate interpretation until exFAT is "well known" and comparable expertise as with FAT16 and FAT32 is achieved. After a detailed description of exFAT structures, the author explained how it would be possible to recover a deleted fragmented file by using the remaining FAT data but did not cover common events, like file move or renaming and in what way bitmap analysis may improve the forensic interpretation accuracy.

Two years later, based on Shullich work, an article [10] on how exFAT may improve file system efficiency and reliability was published. It described and proposed to enhance exFAT structure but did not tackle any forensic uses.

In 2015, Ma, Wang and Cheng [11] described how to improve the reconstruction of deleted file on exFAT systems, based on file characteristics and file combined with statistical analysis. Although focusing on file reconstruction, with excellent statistical results when compared with some digital forensic software, the research does not provide ways to interpret unused entries and file status or metadata artifacts.

Recently, several researches provided an enhanced carving methodology [13][14], focusing on file reconstruction. The carving uses mainly files properties to improve the recovery, as proposed by Uzun and Sencar [15] on JPEG files fragments and developed in some detailed process, like by De Bock and De Smedt [16] on how to automatically recover deleted files.

Alhussein et al. proposed FFS_exFAT [17], a modified fuse exFAT driver to improve fragmented file recovery. However, the proposed solution does not meet exFAT implementation specifications encountered on real world volumes to be forensically analysed in criminal cases investigations.

Moreover, all the detailed description of exFAT structure and carving processes are mainly based on reverse engineering research. Existing research in the literature do not provide clear answers to questions raised by a forensic analyst. These questions are as follows:

- How to make the difference between a file deleted, moved or renamed?
- How to improve findings interpretation accuracy with existing information embedded in system files?
- How to relate carved files with remaining directory inactive entries and associated metadata?

Without any answer to these questions, forensic practitioner's reports will lack soundness, keeping file history vague and unexplained. Although such approach may be welcome in a data recovery process, it is obvious that it will not be acceptable as court evidence. This issue was already raised in 2004 by Buchholz and Spafford [18], explaining clearly the relevance of metadata for forensics to explain:

- who performs action on files?
- where files are coming from?
- when and how it did happen?
- what was done on the file?

In order to address these questions, we follow a structured research approach providing an innovative methodology.

## 3. EXFAT STRUCTURE

### 3.1 FAT before exFAT

Former popular FAT16 and FAT32 file systems were largely covered in previous researches. Carrier [8] provided a clear explanation about the dual-purpose of FAT in terms of cluster allocation and chaining. To facilitate further reading of this paper, we will recap some basic FAT concepts.

Each FAT starts with 8 (FAT32) or 4 (FAT16) bytes used by the system to save other information. Next bytes, grouped in FAT cells, represent all volume clusters starting with cluster 2 as first cluster on a FAT volume is always numbered 2. While FAT16 uses the 2 bytes (16 bits) cells, FAT32 cells uses 4 bytes (but only 28 bits of it). All values are stored using little endian encoding.

- FAT cluster allocation role

Each cell represents a cluster. The cluster is considered as unallocated if the cell is full of 0x00. Any other cell value other than 0x00 means that the cluster is allocated or considered as bad.

- FAT cluster chaining role

The first cluster of each file is saved in the associated directory entry. The value saved in the FAT indicates whether the cluster is the only one used by the file 0xFFFF for FAT16 and 0x0FFFFFFF for FAT32 (all possible bits sets on 1). If more clusters are used by the file, the value of each FAT cells points to the next cluster number, thereby creating a "chain". In FAT16 and FAT32, chaining is used for all files, ignoring whether the file is fragmented or contiguous. A straight consequence of the chaining indicates that associated clusters are flagged as allocated.

When a file is deleted, all FAT cells associated with the formerly used clusters are replaced with 0x00, flagging the associated cluster as unallocated. With common FAT file system drivers, the deleted file entry is flagged by changing the first byte, but the first cluster and the file size are still available until they are overwritten by a new file entry. This behaviour allows to recover contiguous deleted files and at least the first segment of fragmented files.

### 3.2 General structure of an exFAT volume

Shullich [9] provided a detailed description of the exFAT structure. We only highlight the features related to this work.

exFAT, like any file system, provides key information in the Volume Boot Record (VBR). VBR is located at sector #0 of the volume.

Unlike former FAT systems working with two twin FATs, exFAT makes use of only one FAT, still located in the system area (before cluster heap). The offset of the Root Directory can be then localized with the following formula:

ROOTDIRoffset = HEAPs + (ROOTc * Csize), where ROOTDIRoffset is the offset in sectors of the Root Directory relative to the volume start, HEAPs is the amount of sectors reserved for the "system" zone, ROOTc the first cluster allocated to the Root Directory and CSize the amount of sectors in a cluster. An additional useful parameter is provided by the VBR: SSize, sector size, usually set to 512 bytes.

## 3.3 FAT area of exFAT filesystem

The FAT area starts at the sector specified at offset 0x50 in the VBR. Like FAT32, exFAT clusters chaining values are saved in 4 bytes (double-word little endian) cells. However, FAT32 uses only 28 of the 32 bits, exFAT uses the full 32-bits range.

A notable difference with FAT32 is that the FAT area is no more used for allocation, as the BITMAP file takes care of it, although FAT is still used for cluster chaining in case of fragmented files. Additionally, system files like Root Directory and BITMAP file are always flagged as allocated into the FAT too. For all other files the driver should avoid using FAT area whenever possible.

When a file is NOT fragmented, the first cluster number (Cn) and file size in bytes (FILEsize) saved in the file entry, provide enough information to allow a driver to load the file content by reading FILEsize bytes starting from the first cluster.

As a consequence of this new FAT handling, a FAT cluster cell does not provide information about associated cluster allocation. When a FAT cell is filled with 0x00, the associated cluster can be either unallocated or allocated to a non-fragmented file. Moreover, we will see later that even if it contains some value, it is possible that the associated cluster is unallocated. The cluster allocation is now fully handled by the BITMAP and the FAT is dedicated to a single task: cluster chaining.

## 3.4 Root Directory

Like FAT16 and FAT32, the exFAT root directory and all sub-directories are made of sequences of 32-byte entries. However, that is where the similarity ends. The exFAT directory entries structure is totally different from the former FAT implementation. Beside the regular files and folders entries, the Root directory contains unique 32-byte entries. The volume label entry starts with 0x83, the Bitmap starts with 0x81 and the Uppercase table starts with 0x82.

## 3.5 Bitmap file

The Bitmap file, unnamed but referenced in the Root Directory, is identified by its entry first byte (0x81). This entry provides the first cluster (double-word (Dword) at 0x14) and the file size in bytes (two double-word (Qword) at 0x18). Within the BITMAP file, each cluster allocation is flagged by a single bit. The first byte of the BITMAP file represents the 8 first clusters, starting with cluster 2 (the first real cluster). Cluster 2 allocation flag is saved in the least significant bits and Cluster 9 in the most significant bits. The flag of the allocation status of Cluster 10 flag is saved in the least significant bits of the second byte. Knowing that the allocation status in the BITMAP file starts with the status of cluster 2, the flag position of any cluster Cn within that BITMAP file can be computed by using the following formulas:

Byte position value = int((Cn – 2)/8)

Bit position within that byte = Cn – (8 * int((Cn-2)/8))

## 3.6 Regular files and folders entries

Each file or folder is described in its parent folder by several 32-byte entries. The first byte of the entry is used to determine what the entry is describing.

A directory entry, (Figure 1, highlighted in red), starts with 0x85. The next byte describes how many sub-entries are part of the set, followed by a hash value of the file name, DOS attributes and date-time metadata about the file creation, modification and last access.

```
85 02 19 E0 20 00 00 00   27 4B 3E 49 FD 65 86 46
27 4B 3E 49 99 00 88 88   88 00 00 00 00 00 00 00
C0 03 00 0A B0 42 00 00   8B 95 0E 00 00 00 00 00
00 00 00 00 0B 00 00 00   8B 95 0E 00 00 00 00 00
C1 00 63 00 6F 00 6C 00   6F 00 72 00 73 00 2E 00
6A 00 70 00 67 00 00 00   00 00 00 00 00 00 00 00
```

Figure 1. Sample set of exFAT entries for a single file.

First entry is to be considered as the "main" entry describing a file or a folder.

| Offset | Length  | Value | Description               |
|--------|---------|-------|---------------------------|
| 0x00   | 1 byte  | 0x85  | Entry type descriptor     |
| 0x01   | 1 byte  |       | Set entries count         |
| 0x02   | 2 bytes |       | File name hash sum        |
| 0x04   | 1 byte  |       | Dos attributes flags      |
| 0x08   | 4 bytes |       | DOS Date-Time creation    |
| 0x10   | 4 bytes |       | DOS Date-Time modification|

Table 1. Directory entry structure (partial)

| Offset | Length  | Value    | Description                |
|--------|---------|----------|----------------------------|
| 0x00   | 1 byte  | 0xC0     | Entry type descriptor      |
| 0x01   | 1 byte  |          | Flags (at bit #1 FAT use)  |
| 0x03   | 1 byte  |          | File name length           |
| 0x14   | 4 bytes | Cn       | First cluster number       |
| 0x18   | 8 bytes | FILEsize | File size in bytes         |

Table 2. Allocation extension structure (partial)

The Allocation extension, (Figure 1, highlighted in green) follows straight the Directory entry and starts with 0xC0. This entry provides information about file allocation.

The first cluster number (Dword at 0x14) and the file size (FILEsize) in bytes (Qword at 0x18) are self-explanatory. The second bit from second byte is a flag about FAT usage for that file. If the bit is set, it means that FAT is not used for the cluster chaining. It also means that the file is not fragmented and can be retrieved by loading FILEsize bytes from the first cluster position. If the bit is not set, it means that FAT is used for cluster chaining. The file then needs to be loaded by using FAT chaining as for FAT32. We will explain later how the same process may also allow to recover deleted files. During this research while using Windows 10 driver we only encountered an unset exFAT usage bit when the files were really fragmented. At stage 5 of our forensic experiments (cf. Section 4), we observed that when a fragmented file is shrunk to fit in only one fragment, flag in bit 2 of the second byte is set, meaning that FAT is not used anymore.

Because the FAT cells were not updated, one can try to forensically rebuild the former FAT chaining starting from the first cluster and use it to get previous content of the file. We will explain later how to improve the interpretation of the results. This allocation extension provides, at offset 0x03, a single byte with the file name length. All file names are obviously limited to 255 Unicode characters.

File name extension(s): One or several extensions will follow the allocation extension, depending of the file name length. All these extensions start with 0xC1 and contain a maximum of 15 Unicode characters each. In our sample, a simple file name "colors.jpg" is highlighted in yellow (Figure 1).

### 3.7 Inactive entries

There are several reasons for an entry, or a set of entries to become "inactive". This is related with exFAT driver minimizing writings as much as possible.

An entry is set as inactive by un-setting the most significant bit of the entry, changing the first byte original value 0x85, 0xC0 or 0xC1 into 0x05, 0x40 or 0x41, respectively.

In our experiments (Section 5), we observed and verified that when a filename is modified to get a longer name that does not fit anymore into the existing available 0xC1 associated extensions, the driver will set all existing entries related to the file (0x85, 0xC0 and 0xC1's) as inactive. A fully new set of entries will be saved at the end of the folder's existing entries. This avoids reorganizing the whole folder by moving all other entries and limits the update process to the replacement of the first byte of each former entry and the creation of the new entry set. This allows some highly interesting forensic interpretations about the file renaming. Former file name may be found, as it shares unmodified information with the new set of entries: the same first cluster, date-time information and file size.

When a file is moved to another folder in the same exFAT volume, the associated set of entries in the former folder are simply flagged as inactive and an entire set is created into the new folder. Again, the same forensic interpretation can be carried out to identify the location from where the existing file was moved by comparing "active" entries with "inactive" ones in other folders.

In our experiments, we observed that when a file is deleted all related entries and extensions are set as inactive. All the bits in the Bitmap file associated with the clusters and previously allocated to that file are set to 0.

As a consequence, the interpretation that a file was deleted cannot only be based on the fact that the entry is inactive: it is possible that the file was renamed or moved.

### 4. FORENSIC EXPERIMENTATION

In this section, we describe how we reverse engineered the exFAT file structures that were created and modified by Windows native exFAT.sys driver.

In this research, we consider a file as not fragmented if its content is sequentially stored within a set of contiguous clusters. A file is considered as fragmented if its content is stored in at least two non-contiguous clusters.

The created file structure includes files among which are renamed, moved or deleted. We applied the following protocol in order to check the impact of each action at the file level and updates on entries.

- Stage 1: Volume creation and formatting

We create an empty Windows 10 VHD virtual volume file. The created volume is then formatted by using the command line "format" utility specifying exFAT as the file system and with the cluster size of 1024 bytes. The partition was named at the formatting time. The short file name creation mechanism is disabled by default.

- Stage 2: Adding files and folder

Several JPEG and PDF files with different sizes are copied on the volume and a single folder named "subfolder" is created. To facilitate later fragmented files creation, the empty clusters space is then filled by adding dummy files full of zeroes.

- Stage 3: Adding fragmented files

The non-contiguous small files and dummy files, are deleted from the command line and two files were added to occupy previously freed fragmented area. As there are not enough contiguous clusters, both files are fragmented.

- Stage 4: File renaming and moving

Two files are renamed with a longer name (i.e. longer than 16 chars) and two other files are moved to the subfolder.

- Stage 5: File content shortening

The existing fragmented text file added at Stage 3 was shortened so that the total file size will be able to fit into contiguous clusters originally allocated to this file.

- Stage 6: File deletion

Several files, including non-fragmented and fragmented files, were deleted, using the "del" command line utility.

- Stage 7: Subfolder is deleted

To guarantee artefacts reproducibility and mimic the real driver behaviour, the volume is unmounted after each stage and the exFAT partition extracted by using the "dd" Linux command line utility. The resulting exported file is then set in read-only mode. The low level forensic analysis in the dd raw files was performed at byte level by using their own forensic tool [12], combined with the xxd command line Linux hexadecimal viewer. Some scripts were developed to facilitate a cross-check of findings to toggle between the FAT associated word value and the BITMAP file associated bit. The observed artefacts and drivers inhabits were then cross-checked by generating similar set of files and folders on different devices with different partition sizes.

## 5. REBUILD DELETE FILES METHODOLOGY

Based on the observed modification on the directory entries, BITMAP file and FAT cells during the experiments, we propose a methodology to improve interpretation of observable artefacts.

When a file is deleted, two different cases are possible:

• The file did not use FAT chaining (bit set). In this case, the file is not fragmented, and the file content can be recovered by using the available information on the first cluster and FILEsize. The recovery will then be done by saving all bytes starting from the first cluster as referenced into the allocation extension directory entry. Compared with usual carving features, the proposed methodology preserve associated metadata and guarantee that whole file content is recovered.

• The file did use FAT chaining. While this is not possible in FAT16 and FAT32, it is possible in exFAT as, by deleting the file, the file system driver will only update associated clusters in the Bitmap allocation file and avoids updating the FAT to minimize unnecessary writings on the device. In this case, the cluster chain must be rebuilt and list all concatenated clusters content. The last cluster content needs to be truncated in accordance with file size.

```
05 03 04 56 20 00 00 00    33 94 45 49 FB 5B 17 49
33 94 45 49 9B 00 88 88    88 00 00 00 00 00 00 00
40 01 00 10 64 08 00 00    73 A2 56 00 00 00 00 00
00 00 00 00 F5 24 00 00    73 A2 56 00 00 00 00 00
41 00 74 00 61 00 72 00    67 00 65 00 74 00 5F 00
65 00 61 00 72 00 74 00    68 00 2E 00 70 00 6E 00
41 00 67 00 00 00 00 00    00 00 00 00 00 00 00 00
00 00 00 00 00 00 00 00    00 00 00 00 00 00 00 00
```

Figure 2. Set of inactive directory entries describing a file making use of the FAT

In the example in Figure 2, the allocation entry is flagged as using FAT chaining, and the file content starts at cluster 9461 (0x24F5).

Figure 3. Associated FAT cells

Looking into the FAT at associated FAT 4-byte cell, with relative offset 37844 chaining is still available (Figure 3), describing next clusters as 0x24F6, 0x24F7 and beyond.

The filename and the file system metadata can be recovered, as well as date-time values from the Directory entry and the first cluster and file size from the available extensions.

There is, however, an assumption that the file is effectively deleted and that the content is not overwritten. If in most criminal cases the recovered content is obvious, in some circumstances a more robust methodology must be applied, for example when recovering and analysing huge log files.

### 5.1. Validation of the file recovery process

Based on exFAT file system properties, and considering how drivers try to minimize writings whenever possible, some cross-checks need to be done to allow an accurate interpretation of the file status and history:

- Checking the cluster allocation state in the Bitmap file: The list of clusters included in the recovery process needs to be checked within the Bitmap file. If one cluster state is allocated (bit set), then this has to be reported. Such situation may occur when a deleted file was partially overwritten by a more recent one. Depending of the nature of the recovered file, overwritten or reallocated clusters can be replaced by dummy ones, full of null bytes, or simply ignored.
- Checking the cluster chaining in the FAT area: If the file was fragmented (let's call it file1), the cluster chain rebuild needs to be done by the use of the remaining FAT cells. However, a sound analysis of the FAT chaining needs to be carried out to identify potential reuse of the cluster by a more recent file, also being deleted. Identified FILEsize in combination with file content coherence may help during this process.
- Checking the files with similar properties: To avoid wrong interpretation of "deletion" of file, a check must be done in the folder and all other folders to detect potentially renamed or moved file.

Figure 4 Inactive entry

Some helpful properties are the file first cluster, the file size, and the file name. Although all these properties are helpful, a double check will often be needed.

The set of inactive entries in Figure 4 describes a file with the first cluster at 0x02F7 (from the little endian Dword at offset 0x14 of the allocation entry) for a 0x4BAA68 FILEsize.

The file content always starts at the same offset. The first cluster is the most trustable information to be used for similar properties. The file size property is subject to its modification after the move. In the last case, a deeper analysis on date-time values needs to be done, by comparing the creation date-time (should be identical) and the modification date-time (new location would be the most recent one).

Figure. 5 Matching active entry in another folder

The file name can be checked through all the volume using the file name extensions or, to speed up the process, the file name checksum word stored at Offset 0x02 of the directory entry. Although this quick check can provide a good indication, it is still possible to have more files with the same name in different folders. In the case of file renaming, checking the file extension will provide an additional confirmation, if needed.

### 5.2. Carving process

Same exFAT properties may be used to define a simple and efficient carving process. Instead of starting from the inactive directory entries analysis, we propose to start from the bitmap file analysis.

As the deleted files clusters are always flagged as unallocated in the bitmap, it is obvious that looking only at unallocated clusters will speed up the carving process and avoid to "recover" existing and allocated files.

Moreover, we observed in our experiments that Windows exFAT driver avoids creating fragmented files. The availability of information from the bitmap allocation file allows to quickly identify where contiguous space is available to save file content. When FILEsize is known by the Operating System API before file writing process it becomes easy for the exFAT driver to identify that a group of unallocated clusters is large enough to save the file un-fragmented.

This file system driver behaviour needed to be addressed in our experiments by adding Stage 2 (cf. Section 4) to allow later creation of fragmented files by fully filling the available volume space and then deleting small non-contiguous files, so forcing to save larger fragmented files.

Considering real world cases, for such exFAT volumes created on SD memory cards in digital cameras, the fragmentation will probably seldom be encountered. However, if encountered, the use of FAT chaining, whenever available, will still improve the carving process.

When files are modified, and file content becomes too large to fit within the already allocated space, additional clusters are allocated to the file and the file will then become fragmented.

- Identifying the file start by file type header: Like regular carving process the first step is to identify files by the "well known" list of file type headers (i.e. 0xFFD8 for JPEG files). When the speed is essential, the option is to limit the search at the beginning of each cluster, or if VBR info is not available for each sector. This approach, based on the fact that a file always starts at the beginning of a cluster, will not carve files embedded in other files. Photorec [19] software already allows to "scan for files from exFAT unallocated space only" and provides the first sector number in the resulting XML report.
- Searching for all directory entries for identified starting cluster: Now that the starting sector is identified from the results of the carving process, the search for this sector number should be carried on all directory entries, in all folders of the volume, including deleted ones. Searching for a pattern starting with byte 0x40 (deleted allocation extension file entry), ignoring 21 indifferent bytes, and checking if next 4 bytes match the starting sector value (little endian). In the case of some tools output (e.g. Photorec), the conversion from Sector to Cluster needs to be computed, by taking the exFAT system area and the cluster size into account. As the bitmap shows that the cluster is unallocated, no existing file would match this pattern. Depending the volume file activity, it is possible that several entries match the pattern.
- Select the most recent directory main entry: If multiple matches are possible, it can be explained by moved or renamed files before deletion or by multiple files overwrite. As the carving will extract content from the most recent file, it will then be important to compare matches and deduce the most recent entry update. Unlike NTFS MFT record updates, exFAT does not provide such information. However, available date-time values (creation, last content modification, and last access) may be used to establish the most recent directory entry. A simple script may scan the volume to gather all sequences of bytes matching exFAT main directory entry pattern. Ideally, such search would be saved in a database to facilitate and speed up multiple files carving.
- Recover file size from allocation entry: Carving tools often use the file embedded metadata, when available, or the existence of file type well know footer (like 0xFFD9 for JPEG files) to identify or compute real file size. We advise to compare carved content size with information available in the identified file associated allocation entry (0x40). In some exceptional cases the file may be fragmented and carved content size will not match FILEsize value. The forensic analyst will then be able to explain why values do not match by showing "do not use FAT" flag bit on "0". Moreover, it will be possible to provide a more accurate file content recovery by using potentially remaining FAT chaining as early demonstrated.
- Document carved file with the available metadata: Previous stages already allowed to recover metadata concerning file size, date and time values. The remaining associated file name entries (0x41) will allow to rebuild, at least partially, carved file name.

### 5.3. Recover former content from existing file

When a file content is shortened, resulting in the lower FILEsize value in the allocation entry, exFAT properties will allow to get "a bit more" content.

When identifying a deleted or existing file related to the case, forensic analyst may look at the bitmap area and FAT cluster chains.

- A bit more from shortened fragmented file: When shortening a fragmented file, the driver will only update FILEsize value, former FAT chaining cells will not be updated. A simple rebuild based on FAT chaining may allow to recover former file content. In this case only clusters flagged as unallocated in bitmap allocation file are to be considered.
- A bit more from shortened non-fragmented file: Another interesting use of bitmap cluster allocation file analysis is to check for whether contiguous clusters to the identified existing file logical end are flagged as unallocated.

Let's suppose that the analysed non-fragmented file starts at cluster 530 with an actual FILEsize requiring use of 2 clusters. Associated bitmap bits for clusters 530 to 537 are:

| 1 | 1 | 0 | 0 | 0 | 0 | 1 | 1 |
|---|---|---|---|---|---|---|---|

Clusters 532 to 535 being unallocated, it would be possible that it contains previous version of the file, when it was larger. This may be validated by checking the next folder file entries. If next file entry shows that the first cluster value is cluster 536, it indicates that this file was created sequentially after the analysed file and that clusters 532 to 535 may be allocated to a former version of the file content.

Before validating the content of the identified clusters as part of a previous version of file content of interest, it is necessary to check if any other inactive entry is not starting with one of these clusters and perform a cross-check by analysing related FAT cells. If any chaining is encountered in the FAT, the start of the chain should match the file's first cluster.

### 6. CONCLUSION AND FUTURE WORK

The methodology proposed in this paper addresses several identified issues and provides a suitable forensic process. Human expertise is required to analyse file systems at byte level. Almost all forensics software tools with hexadecimal viewer and basic search features will facilitate such search and speed the process, especially if scripting features are available. It quickly becomes obvious that, to facilitate forensic analyst examination, there is a need for a software feature allowing to jump from a cluster to the associated bitmap or FAT entries and inversely. Adding this feature to our file system analysis software Tyrhex [12] highly facilitated the analysis process. A four days training on File Systems Forensics was hosted by European Union Agency for Law Enforcement Training in 2017, EU law enforcement members applied our proposed methodology successfully by using basic byte level search features with the popular forensic tools.

If exFAT interpretation provided by the tools needs to be checked, using a basic hexadecimal viewer allows to recover exFAT fragmented file and associated metadata. Participants who recovered a text file partially by using carving were able to rebuild the missing part from next fragmented sequence of clusters and validate the file length by recovering FILEsize value from directory entry artefacts. Integrity of recovered file was then demonstrated by Bitmap analysis. Facing inaccurate information provided by XWays on a JPEG file "moved or renamed" they were able to demonstrate that the file was simply moved to another folder named "under14" on the same volume. For child pornography cases, this additional clue can only be highly significant evidence in court, demonstrating that the suspect knowingly sorted the pictures on subject supposed age. Beyond important validation of recovered file content, the proposed methodology allows to correlate the metadata information found in the remaining inactive directory entries with carved files.

Using the described processes, forensic analysts will be able to explain how forensic software handles file recovery and decide, whenever needed, to improve results by adding human processing.

Future work will aim to test how most popular forensics software tools handle automatic recovery of exFAT fragmented deleted files, interpret accurately files moved or renamed and link with associated metadata. We are also looking at applying our approach in mobile device forensics [23], vehicle forensics [24] and investigation of IoT devices [25].